\documentclass{PoS}
\usepackage{amssymb, fontenc, times, mathptmx, graphicx}
\title{Kinematic differences between NLS1 and BLAGN sources}

\ShortTitle{Kinematic differences between NLS1 and BLAGN sources}

\author{Ma\v{s}a Laki\'{c}evi\'{c}\\ 
        Astronomical Observatory Belgrade; Volgina 7, 11060 Belgrade, Serbia\\
        E-mail: \email{mlakicevic@aob.rs}}

\author{\speaker{Luka \v{C}. Popovi\'{c}}\\
        Astronomical Observatory Belgrade; Volgina 7, 11060 Belgrade, Serbia\\}
        
\author{Jelena Kova\v{c}evi\'{c}-Doj\v{c}inovi\'{c}\\ 
        Astronomical Observatory Belgrade; Volgina 7, 11060 Belgrade, Serbia\\}       

\abstract{It is well-known that the higher policyclic aromatic hydrocarbon (PAH) abundance, lower black hole mass, higher accretion rate and lower luminosities are among the major characteristics of Narrow-Line Seyfert galaxies (NLS1), when they are compared to Broad line Seyfert galaxies (BLS1). NLS1s may be normal Seyfert galaxies at an early stage of evolution, their black holes may still be growing and/or they could be special for some other reason. In this work we discuss the findings that NLS1s have most of line and continuum luminosities correlated with FWHM(H$\beta$), which may be the trace of their rapid black hole mass grow. BLS1 do not show such trends. Also, PAHs may be destroyed as the black hole grows and the starbursts are removed, for NLS1 objects.}

\FullConference{Revisiting narrow-line Seyfert 1 galaxies and their place in the Universe - NLS1 Padova\\
		9-13 April 2018 \\
		Padova Botanical Garden, Italy}

\begin{document}
\section{Introduction}
Narrow-line Seyfert 1 (NLS1) galaxies show relatively narrow full width at half maximum (FWHM$\le$2000 km s$^{-1}$, \cite{Veron01}) of permitted lines, narrower than in typical Seyfert 1 galaxies. NLS1s have certain specific characteristics with respect to the broad line active galactic nuclei (BLAGNs). NLS1 objects have more polycyclic aromatic hydrocarbons (PAHs) \cite{Sani10}, more dust spirals and bars \cite{Deo06} and higher accretion rates \cite{Bian03} than BLAGNs, while BLAGNs have higher black hole (BH) masses (M$_{\rm BH}$) \cite{Mathur00}, and higher optical, X-ray and UV luminosities \cite{Lakicevic18,Grupe10}. NLS1s have lower optical variability \cite{Rakshit17b}, higher X-ray variability \cite{Zhou06} and possibly a lower inclination \cite{Rakshit17a}, than the BLAGNs. NLS1s have strongest Fe\,II emission, strongest X-ray excess, and largest C\,IV blueshifts \cite{Shapovalova12}. It is believed that NLS1s are AGNs in the early stage of evolution \cite{Mathur00,Mathur01} and that their black holes are growing \cite{Jin12a}. 

The aim of this research was to explore if NLS1s show some characteristics different from BLAGNs, using some optical and mid-infrared (MIR) spectroscopic parameters, as this may suggest the nature of differences between these two groups of objects. Here, we present some correlations that are different for NLS1s and BLAGNs. We found that NLS1s show the dependence of most of line and continuum luminosities and PAH with FWHM(H$\beta$), while BLAGNs do not show these trends. 

\section{The sample and method of analysis}
In this work we adopted the criteria for NLS1s that FWHM(H$\beta$)$\le$2200 km s$^{-1}$ and that their flux ratio of total [O III] $\lambda$5007 to total H$\beta$ is <3 \cite{Rakshit17a}. The sample consists of 64 NLS1s and 99 BLAGNs, with redshifts <0.7, that have optical (Sloan Digital Sky Survey -- SDSS DR12) and MIR (InfraRed Spectrograph -- IRS on Spitzer Space Telescope) spectral parameters, described and available in papers \cite{Lakicevic17} and \cite{Lakicevic18}. Spectral parameters are: 
\begin{itemize}
 \item Optical: FWHM(H$\beta$), AGN continuum luminosity at 5100 \AA, (L5100), Luminosity of H$\beta$ broad line\footnote{We fitted the broad spectral line H$\beta$ with three Gaussians: narrow, intermediate and very broad. For FWHM(H$\beta$) and LH$\beta$b we included both intermediate line region and very broad line region.} (LH$\beta$b) and M$_{\rm BH}$.
 \item MIR: fractional contribution of PAH component to the integrated 5--15 $\mu$m luminosity -- RPAH, monochromatic luminosity of the source at 6 $\mu$m -- L6 and the luminosities of coronal MIR lines, [O\,IV] at 25.89 $\mu$m and [Ne\,V] at 14.32 $\mu$m -- L[O\,IV]25.89 and L[Ne\,V]14.32. 
 \end{itemize} RPAH and L6 were calculated using deblendIRS code \cite{Hernan15}. The coronal MIR lines were fitted using the Gaussian function and the luminosities are obtained. 

\section{Results}
\begin{figure}[!tbp]
  \centering
  \begin{minipage}[b]{0.45\textwidth}
    \includegraphics[width=\textwidth]{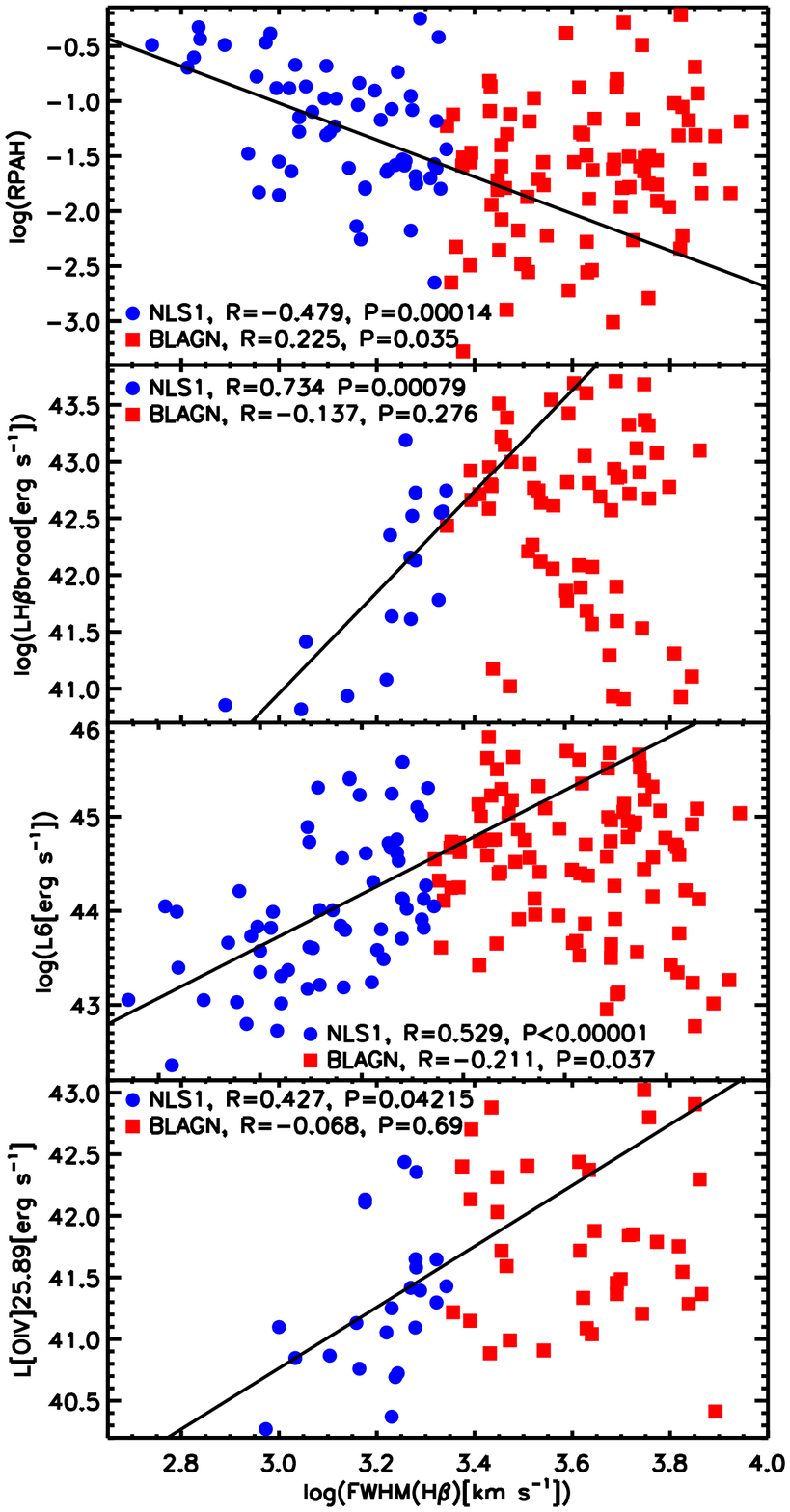}
    \caption{FWHM(H$\beta$) compared to other parameters: RPAH, LH$\beta$b, L6 and L[O\,IV]25.89, for NLS1s and BLAGNs.  \label{fig:prva}}
  \end{minipage}
  \hfill
  \begin{minipage}[b]{0.45\textwidth}
    \includegraphics[width=\textwidth]{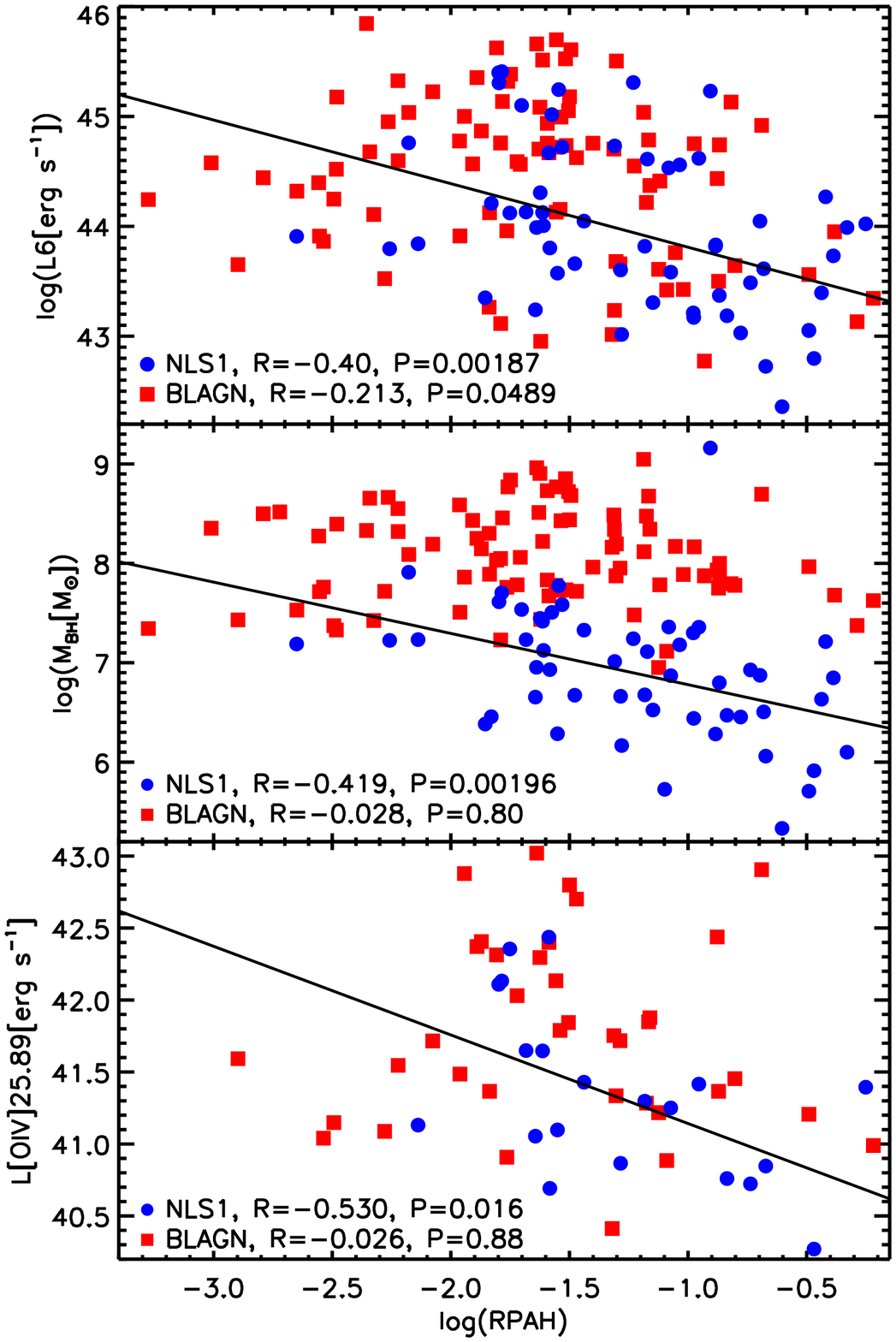}
    \caption{RPAH decreases with L6 (upper panel), with M$_{\rm BH}$ (middle panel) and with L[O\,IV]25.89 (bottom panel), for NLS1 objects. BLAGNs do not show any trends. \label{fig:druga}}
  \end{minipage}
\end{figure}

In Fig.~\ref{fig:prva} we present the FWHM(H$\beta$) vs. several optical and MIR parameters: RPAH, LH$\beta$b, L6 and L[O\,IV]25.89. We noticed that NLS1s show the trends (relations with P-values < 0.05 are adopted as significant), while BLAGNs do not show them.

In Fig.~\ref{fig:druga} we present RPAH vs. parameters: L6, M$_{\rm BH}$ and L[O\,IV]25.89. RPAH decreases with L6 and L[O\,IV]25.89 and with M$_{\rm BH}$s in NLS1 objects. BLAGNs do not show any obvious trends.

\begin{figure}
\includegraphics[width=.5\textwidth]{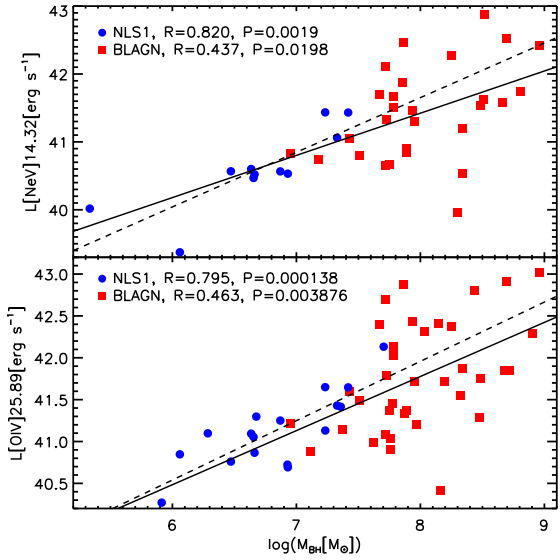}
\caption{Upper panel: correlation L[Ne\,V]14.32--M$_{\rm BH}$; lower panel: correlation L[O\,IV]25.89 -- M$_{\rm BH}$. \label{fig:treca}}
\label{fig:}
\end{figure}

The correlation between L[Ne\,V]14.32 and L[O\,IV]25.89 with M$_{\rm BH}$ was found by \cite{Dasyra08}, who claimed that this correlation is most likely the consequence of the NLR gas kinematics being determined by the potential of the bulge. Thus, these luminosities can be used to estimate M$_{\rm BH}$. We tested this relation in Fig.~\ref{fig:treca}. Significantly stronger trends were obtained for NLS1s than for BLAGNs, for both lines. That may imply that this relation may be more suitable for the estimation of M$_{\rm BH}$ for NLS1 objects than for BLAGNs. 

\section{Discussion}

Regarding the FWHM(H$\beta$)--RPAH trend for NLS1s (Fig.~\ref{fig:prva}), it was already noticed in \cite{Lakicevic17} and references within, but for the whole BLAGN and NLS1 sample. FWHM(H$\beta$)--LH$\beta$b trend (Fig.~\ref{fig:prva}) is found by \cite{Zhou06} and \cite{Veron01}. \cite{Veron01} found it both for NLS1s and BLAGNs. FWHM(H$\beta$) and LH$\beta$b origin from broad line region (BLR), which is gravitationally bound to the BH. Thus, FWHM(H$\beta$) is used to measure the dispersion velocity of that region, while LH$\beta$ most likely also grows with M$_{\rm BH}$. This may be connected with the findings of \cite{Popovic11} that the objects with more starbursts show certain correlations that other Seyferts 1 do not have (such as the correlation FWHM(H$\beta$)--L5100). Finally, the correlations of FWHM(H$\beta$) with L6 and L[O\,IV]25.89, for NLS1s (Fig.~\ref{fig:prva}) could also be explained with BH growth, since 1) FWHM increases with M$_{\rm BH}$ (see above), and 2) luminosities depend on BH size \cite{Sani10}. These trends may be the trace of the evolution of NLS1 objects, as FWHM(H$\beta$), luminosities and M$_{\rm BH}$ grow.  

Anticorrelations of RPAH with luminosities and M$_{\rm BH}$ (Fig.~\ref{fig:druga}) could be understood as the PAH distruction and starburst removal, with BH growth, in NLS1 objects, but also as the natural transition from the type NLS1 to BLAGN type of objects. PAHs may be more actively destroyed in NLS1s than in BLAGNs, as M$_{\rm BH}$ and MIR luminosities grow.

Trends M$_{\rm BH}$--Luminosities of coronal lines (Fig.~\ref{fig:treca}), are found to be considerably higher for NLS1s than for BLAGNs. Perhaps the reason for that is that the MIR narrow line region lines can not well estimate the BLR parameters of BLAGN objects. The future research should examine these trends for the larger samples of objects.

\section{Conclusion}
NLS1s could be the objects with BHs that grow faster than the ones in BLAGNs, as it is proposed in various literature. They have higher accretion rates, more starbursts, but lower line and continuum luminosities at optical and MIR wavelengths and lower M$_{\rm BH}$ than BLAGNs \cite{Lakicevic18}. 
Here we used optical and MIR spectral parameters to explore the correlations that are different between NLS1 and BLAGN objects. These differences are important to understand the nature of these objects. 

The results of our investigations led us to following conclusions:
\begin{itemize}
\item The correlations between FWHM(H$\beta$) and luminosities of MIR and optical lines and continuum, for NLS1s, are probably the consequence of the M$_{\rm BH}$ growth which increases both FWHMs and luminosities. Higher mass increase the luminosities and velocities. 
\item We found that RPAH decreases with M$_{\rm BH}$ and luminosities of coronal lines and continuum, only for NLS1s.
\item The trends M$_{\rm BH}$--Luminosities of coronal lines are higher for NLS1s than for BLAGNs.
\end{itemize}

\section*{Acknowledgements}
This conference has been organized with the support of the Department of Physics and Astronomy ``Galileo Galilei'', the University of Padova, the National Institute of Astrophysics INAF, the Padova Planetarium, and the RadioNet consortium. RadioNet has received funding from the European Union's Horizon 2020 research and innovation programme under grant agreement No~730562. This work was supported by the Ministry of Education, Science and Technological Development of Serbia through the project~176001, "Astrophysical Spectroscopy of Extragalactic Objects".

\end{document}